\documentclass[aps,pra,twocolumn,superscriptaddress]{revtex4}
\usepackage{amsmath}
\usepackage{amsfonts}
\newcommand{\ket}[1]{\left|{#1}\right>}

\DeclareMathOperator*{\minusmin}{--\,min}
\begin{document}

\title{Optimal Lewenstein--Sanpera decomposition of two-qubit states\\ using Semidefinite Programming}


\author{Guo Chuan Thiang}
\affiliation{Centre for Quantum Technologies, National University of Singapore, 3 Science Drive 2, 117543, Singapore}
\affiliation{Department of Physics, National University of Singapore, 2 Science Drive 3, 117542, Singapore}
\author{Philippe Raynal}
\affiliation{Centre for Quantum Technologies, National University of Singapore, 3 Science Drive 2, 117543, Singapore}
\author{Berthold-Georg Englert}
\affiliation{Centre for Quantum Technologies, National University of Singapore, 3 Science Drive 2, 117543, Singapore}
\affiliation{Department of Physics, National University of Singapore, 2 Science Drive 3, 117542, Singapore}

\date{\today}

\begin{abstract}
We use the language of semidefinite programming and duality to derive necessary and sufficient conditions for the optimal Lewenstein--Sanpera Decomposition (LSD) of 2-qubit states. We first provide a simple and natural derivation of the Wellens--Ku\'s equations for full-rank states. Then, we obtain a set of necessary and sufficient conditions for the optimal decomposition of rank-3 states. This closes the gap between the full-rank case, where optimality conditions are given by the Wellens--Ku\'s equations, and the rank-2 case, where the optimal decomposition is analytically known. We also give an analytic expression for the optimal LSD of a special class of rank-3 states. Finally, our formulation ensures efficient numerical procedures to return the optimal LSD for any arbitrary 2-qubit state.
\end{abstract}

\maketitle

\section{Introduction}
Quantum entanglement is perhaps the most striking phenomenon associated with quantum systems. Once seen as ``evidence'' for the alleged incompleteness of quantum mechanics \cite{einstein35}, entanglement has now found numerous applications as a resource for quantum communication \cite{bennett92,bennett93}, computation \cite{shor97}, and cryptography \cite{ekert91}.

The characterization of entanglement has become a key area of quantum information theory. Various schemes to quantify entanglement have been proposed (for a review, see \cite{horodecki09}). A particularly interesting approach is offered by the Lewenstein--Sanpera decomposition (LSD) \cite{lewenstein98} of a composite quantum state, which comprises a convex sum of a separable state and an entangled state. 

Now for any 2-qubit system, there is a unique \emph{optimal} LSD. This optimal decomposition has a separable part with maximal weight, and the entangled part is a pure state. The weight of the pure state in this decomposition multiplied by its concurrence \cite{wootters97,wootters98} provides a measure of entanglement for the 2-qubit state \cite{lewenstein98, wellens01}.

Analytical expressions for the optimal LSD of some special cases were found in \cite{englert00}, these include the rank-2 states, the self-transposed states, and the generalized Werner states. Recently, a pair of coupled nonlinear equations for finding the optimal LSD of full-rank states was obtained by Wellens and Ku\'s \cite{wellens01}. However, an analytic solution to these equations is only available in the case where the separable part in the optimal LSD has full rank. 

As noticed in \cite{Rezaee06}, the problem of finding the optimal LSD can be in some cases formulated as a SemiDefinite Program (SDP). In the present paper, we systematically exploit this connection for 2-qubit states. We first rederive the Wellens--Ku\'s equations for full-rank states in a particularly transparent manner. The SDP formulation also enables us to efficiently compute the optimal decomposition by numerical means. We then extend our analysis to rank-3 states, and obtain necessary and sufficient optimality conditions. With the optimal LSDs of rank-2 states already known \cite{englert02}, this completes the characterization of optimal LSDs for 2-qubit states. We also obtain analytically the optimal LSD for the class of rank-3 states that are orthogonal to a product state and have a separable part of rank 3. For such states, the pure state in the optimal LSD is maximally entangled. This is similar to the full-rank case where the separable part is full rank. There, the nonseparable pure state is maximally entangled too \cite{karnas01}.

\section{Lewenstein--Sanpera Decompositions}

The construction of LSDs hinges on the fact that the set of separable states is convex. Any composite system can be written as a convex sum of a separable state $\rho_\text{sep}$ and an entangled state $\rho_\text{ent}$. Information about nonseparability is then contained in $\rho_\text{ent}$; for example, the state $\rho$ is nonseparable if $\rho_\text{ent}$ does not vanish, and only then.

A simple dimensional argument \cite{lewenstein98} leads to the important consequence that for 2-qubit states, $\rho_\text{ent}$ is just a pure state. In general, there is a continuum of LSDs, $\rho = \lambda\rho_\text{sep}+(1-\lambda)\rho_\text{pure}$, for a given state. Among these is the \emph{optimal} LSD,
\begin{equation}
	\rho = \mathcal{S}\varrho_\text{sep}+(1-\mathcal{S})\varrho_\text{pure},\quad \mathcal{S}=\text{max}\{\lambda\}\label{optlsd},
\end{equation}
where $\mathcal{S}$ is the \emph{degree of separability} of $\rho$. Throughout this paper, we will use calligraphic font to refer to quantities that are optimal. 

When $\rho$ has full rank, $\varrho_\text{sep}$ is either full-rank or \mbox{rank-3}. In the latter situation, we denote its null eigenstate by $\rho_1$. Let us also introduce $\varrho^{\text{T}_1}_\text{sep}$, the partial transpose with respect to the first qubit of $\varrho_\text{sep}$. Then the \emph{barely-separable} property of $\varrho_\text{sep}$ \cite{karnas01} says that $\varrho^{\text{T}_1}_\text{sep}$ has a zero eigenvalue, whose corresponding null eigenstate shall be denoted by $\rho_2$. We quote the following results from the Wellens--Ku\'s paper \cite{wellens01}, with slight modifications to their notation.

In the optimal LSD of a full-rank state, $\varrho_\text{pure}$ is an eigenstate of $\mu\rho_1+\rho_2^{\text{T}_1}$, $\mu\geq0$ , with a nonpositive eigenvalue,
\begin{equation}
	\exists\alpha,\mu\geq 0\qquad(\mu\rho_1+\rho_2^{\text{T}_1})\varrho_\text{pure}=-\alpha\varrho_\text{pure},\label{sepred:eig1}
\end{equation}
with $\mu\rho_1 \equiv 0$ if $\varrho_\text{sep}$ has full rank.
This is accompanied by the eigenstate equation for $\rho_2$,
\begin{equation}
	\bigl(\rho-(1-\mathcal{S})\varrho_\text{pure}\bigr)^{\text{T}_1}\rho_2=0.\label{sepred:eig4}
\end{equation}
Equations \eqref{sepred:eig1} and \eqref{sepred:eig4} are the Wellens--Ku\'s equations. In general, there may be several solutions to these coupled eigenvalue equations. However, consistent with the uniqueness of the optimal LSD, there is only one with $\mu,\alpha\geq 0$ that gives a positive and separable, and thus permissible, $\rho_\text{sep}$. 

The original proofs of these assertions, as well as the sufficiency of these equations, involve considerable technical detail. The aim of the present paper is to present an alternative derivation, and to generalize these equations to the reduced-rank case.

\section{Semidefinite Programming}
In semidefinite programming \cite{vandenberghe96}, a linear objective function is minimized subject to the constraint that an affine combination of hermitian matrices is positive semidefinite. We now briefly review some important features of SDP.

\subsection{The primal semidefinite program}
In its canonical form, the primal semidefinite program is formally stated as:
\begin{equation}
	\begin{array}{ll}
	\text{minimize} & \vec{c}^{\,\text{T}}\vec{x}\\
	\text{subject to} & F(\vec{x})\geq 0,\label{primalprogram}
	\end{array}
\end{equation}
where	$F(\vec{x})=F_0+\sum_{i=1}^mx_iF_i$ and $\vec{x}\in\mathbb{R}^m$. The inputs for the primal problem are (i) the vector $\vec{c}\in\mathbb{R}^m$ characterizing the objective function, and (ii) the $m+1$ hermitian matrices $F_0,F_1,\ldots,F_m\in\mathcal{H}^n$ defining the linear matrix inequality, where $\mathcal{H}^n$ is the space of $n\times n$ Hermitian matrices. The primal problem is strictly feasible if there exists $\vec{x}$ such that $F(\vec{x})>0$. The primal optimal value is $p^*=\inf\{\vec{c}^{\,\text{T}}\vec{x}\;|\;F(\vec{x})\geq 0\}$, and we denote the primal optimal set by
\begin{equation}
	\mathbf{X}_\text{opt}=\{\vec{x}\:|\:F(\vec{x})\geq 0\:\, \text{and}\:\, \vec{c}^{\,\text{T}}\vec{x}=p^*\}.\label{primaloptset}
\end{equation}

\subsection{The dual semidefinite program}
The dual problem associated with \eqref{primalprogram} is
\begin{equation}
	\begin{array}{ll}
		\text{maximize} &-\text{tr}\{F_0Z\}\\
		\text{subject to} &\text{tr}\{F_iZ\}=c_i, i=1,\ldots,m,\\
		{} & Z\geq 0.
	\end{array}\label{dualprogram}
\end{equation}
The dual variable $Z=Z^{\dag}\in\mathcal{H}_+^n$ is subject to $m$ equality constraints, defined by the $F_i$s and $c_i$s specified in the primal program, in addition to a condition of nonnegativity. The dual problem is strictly feasible if there exists $Z>0$ satisfying the dual constraints. The dual optimal value is $d^*=\sup\big\{-\text{tr}\{F_0Z\}\;|\;Z\geq 0,\:\text{tr}\{F_iZ\}=c_i\: \forall i\big\}$, while the dual optimal set is
\begin{equation}
	\mathbf{Z}_\text{opt}=\big\{Z\geq0\:|\:\text{tr}\{F_iZ\}=c_i\: \forall i, -\text{tr}\{F_0Z\}=d^*\big\}\label{dualoptset}.
\end{equation}
One also has the hierarchy $-\text{tr}\{F_0Z\}\leq d^*\leq p^*\leq \vec{c}^{\,\text{T}}\vec{x}$, meaning that the dual objective yields lower bounds on the optimal primal value, while the primal objective yields upper bounds on the optimal dual value.

\subsection{Complementary slackness condition}
An important quantity to consider is the duality gap $\vec{c}^{\text{T}}\vec{x}+\text{tr}\{F_0Z\}=\text{tr}\{F(\vec{x})Z\}$, which is a nonnegative quantity linear in $\vec{x}$ and $Z$. The equality $d^*=p^*$ holds (no duality gap) if either the primal or the dual problem is strictly feasible. If \emph{both} are strictly feasible, the optimal sets $\mathbf{X}_\text{opt}$ and $\mathbf{Z}_\text{opt}$ are nonempty, and there exist feasible pairs of $\vec{x}$ and $Z$ with $p^*=\vec{c}^{\,\text{T}}\vec{x}=-\text{tr}\{F_0Z\}=d^*$, so that $F(\vec{x})Z=0$. This is the \emph{complementary slackness condition}, stating that the ranges of the nonnegative matrices $F(\vec{x})$ and $Z$ are orthogonal. Under strict primal and dual feasibility, one then has necessary and sufficient optimality conditions for the semidefinite program: a feasible $\vec{x}$ is optimal if and only if there exists a $Z$ such that
\begin{equation}
	\begin{array}{l}
		F(\vec{x})\geq 0,\: Z\geq 0,\\
		\text{tr}\{F_iZ\}=c_i,\: i=1,\ldots,m,\\
		F(\vec{x})Z=0.
	\end{array}\label{optcondition}
\end{equation}
The above equations provide algebraic expressions that the optimal $\vec{x}$ and $Z$ must satisfy. We will see in the next section that these conditions lead to the Wellens--Ku\'s equations.

\section{Derivation of the Wellens--Ku\'s equations}
Let us notice that we have an optimization problem, in which we wish to minimize a scalar function $-\lambda=-\text{tr}\{\lambda\rho_\text{sep}\}$ of some variables subject to a set of constraints. Firstly, we require $\rho_\text{sep}$ and $\rho_\text{pure}$ in a LSD to be positive semidefinite. Next, the Peres--Horodecki criterion \cite{peres96,horodecki96} tells us that a 2-qubit state is separable if and only if its partial transpose is positive. The crucial point here is that the \emph{separability} constraint has become a \emph{positivity} constraint, ensuring that the optimal LSD problem for 2-qubit states can be formulated as a SDP. We will proceed to show this explicitly. For simplicity, we only consider full-rank states in this section. The case of reduced-rank states will be considered in the following section.

\subsection{Optimal LSD as a semidefinite program}
\subsubsection{The primal problem}
We use $\vec{\sigma}$ and $\vec{\tau}$ to denote the Pauli operators in the first and second qubit space, respectively. It will be convenient to use the \emph{magic basis}, introduced by Hill and Wootters \cite{wootters97, wootters98}, in which the Pauli operators are represented by imaginary antisymmetric $4 \times 4$ matrices while their products are represented by real, symmetric matrices. Partial transposition in the first qubit is effected by $\vec{\sigma}\to-\vec{\sigma}, \vec{\tau}\to\vec{\tau}$.

Our basis $\{E_i : i=1,\ldots,16\}$ for $4\times 4$ hermitian operators comprises the sixteen combinations of the Pauli operators and the identity, $\sigma_i\tau_j$, where $i,j=0,1,2,3$ and $\sigma_0=\tau_0=\openone_4$. These are traceless (except $E_1=\openone_4$) and mutually orthogonal, i.e., $\text{tr}\{E_iE_j\}=4\delta_{ij}$. A LSD of a state $\rho$ can be written as
\begin{equation}
\rho=\lambda\rho_\text{sep}+(1-\lambda)\rho_\text{pure}\equiv\tilde{\rho}_\text{sep}+\tilde{\rho}_\text{pure},
\end{equation}
where the weights $\lambda$ and $1-\lambda$ have been absorbed into $\tilde{\rho}_\text{sep}\equiv\lambda\rho_\text{sep}$ and ${\tilde{\rho}_\text{pure}\equiv(1-\lambda)\rho_\text{pure}}$. In this notation, we have the parameterization $\tilde{\rho}_\text{sep}=\frac{1}{4}\vec{x}\cdot\vec{E}$, where ${\vec{x}^{\,\text{T}}=(\lambda,x_2,\ldots,x_{16})\in\mathbb{R}^{16}}$.

In the search for the optimal LSD, we comb through the possible $\tilde{\rho}_\text{sep}$s via choices of $\vec{x}$, but these choices are not arbitrary. To ensure a valid decomposition in the first place, we must enforce three constraints,
\begin{equation}
\begin{array}{ll}
	\begin{tabular}[c]{ll}
	(i) & positivity of $\tilde{\rho}_\text{sep}$ \\
	(ii) & separability of $\tilde{\rho}_\text{sep}$ \\
	(iii) & positivity of $\tilde{\rho}_\text{pure}$ \\
	\end{tabular}
\end{array}
\end{equation}
which we merge into a single inequality of a $12 \times 12$ matrix:
\begin{equation}
\begin{array}{ll}
	\begin{bmatrix}
		\tilde{\rho}_\text{sep} & 0 & 0\\
		0 & \tilde{\rho}_\text{sep}^{\text{T}_1} & 0\\
		0 & 0 & \rho-\tilde{\rho}_\text{sep}
	\end{bmatrix}\geq 0.\label{blockconstraint}
\end{array}
\end{equation}
Next, we introduce 16 block-diagonal $12\times12$ hermitian matrices $F_i$ associated with the $E_i$s, defined by $F_i=\frac{1}{4}\text{diag}(E_i, E_i^{\text{T}_1}, -E_i), i=1,\ldots,16$, as well as ${F_0=\text{diag}(0, 0, \rho)}$. In terms of the $F_i$s, the inequality constraint in Eq.~\eqref{blockconstraint} can be expressed as ${F(\vec{x})=F_0+\sum_{i=1}^{16}{x_iF_i}\geq 0}$. Finally, let ${\vec{c}^{\,\text{T}}=(-1, 0,\ldots,0)\in \mathbb{R}^{16}}$, so that ${\vec{c}^{\,\text{T}}\vec{x}=-\lambda}$. Maximizing $\lambda$ to obtain the optimal LSD is then equivalent to minimizing $\vec{c}^{\,\text{T}}\vec{x}$.

With these specifications, we have rephrased the optimal LSD problem as a SDP in the form of \eqref{primalprogram}. One can then efficiently compute the optimal LSD of a given 2-qubit state using well-established algorithms for solving SDPs. For instance, we have written a working routine using \textsf{cvx} version 1.2 \cite{grant08a}, which is a modeling system for disciplined convex programming, utilizing the open-source solver SDPT3 \cite{TTT06}. 

Next, we establish strict primal feasibility. For this, we choose $\tilde{\rho}_\text{sep}=\alpha\frac{1}{4}\openone_4$, i.e., $\vec{x}^{\,\text{T}}=(\alpha, 0,\ldots,0)$, where $\alpha\frac{1}{4}$ is a positive number smaller than the smallest eigenvalue of $\rho$. Clearly, $\tilde{\rho}_\text{sep}>0$ and	$\tilde{\rho}_\text{sep}^{\text{T}_1}>0$. Furthermore, since $\rho>0$, it has a spectrum with 4 positive eigenvalues, and choosing $\alpha$ as described above, the difference ${\rho-\tilde{\rho}_\text{sep}=\rho-\alpha\frac{1}{4}\openone_4}$ is still positive definite. Thus, Eq.~\eqref{blockconstraint} holds with strict inequality as required. Since the primal problem is strictly feasible, we conclude that there is no duality gap.

\subsubsection{The dual problem}
We now focus our attention on the dual problem associated with \eqref{primalprogram}. Following \eqref{dualprogram}, the dual variable $Z$ is a $12\times 12$ positive semidefinite matrix subject to the 16 dual constraints $\text{tr}\{F_iZ\}=c_i$. Since $F_0$ and $F_i$ are block-diagonal, the dual objective depends only on the block-diagonal entries of $Z$. Without loss of generality, we can choose $Z$ to be block-diagonal. For convenience, we write
\begin{equation}
Z=
	\begin{bmatrix}
		Z_1 & 0 & 0\\
		0 & Z_2 & 0\\
		0 & 0 & Z_3
	\end{bmatrix},
\end{equation}
where $Z_1$, $Z_2$ and $Z_3$ are nonnegative $4\times4$ matrices. With this notation, the dual objective becomes $-\text{tr}\{\rho Z_3\}$. Since there is no duality gap, we have ${d^*=-\text{tr}\{\rho\mathcal{Z}_3\}=p^*=-\mathcal{S}}$.

The dual problem is strictly feasible too: choose ${Z=\text{diag}(\openone_4,\openone_4,3\openone_4)>0}$, and check that all the constraints are indeed fulfilled. The first dual constraint ${\text{tr}\{F_1Z\}=-1}$ is satisfied and the $2\text{nd}$ to	$16\text{th}$ dual constraints $\text{tr}\{F_iZ\}=0$ hold, since $E_i$ and $E_i^{\text{T}_1}$ are traceless by construction.
\subsection{Equivalence of complementary slackness condition and Wellens--Ku\'s equations}
With strict primal \emph{and} dual feasibility, we now have necessary and sufficient optimality conditions as a consequence of the complementary slackness condition \eqref{optcondition}. In the present context, conditions \eqref{optcondition} translate into the following statement. The primal variable $\tilde{\varrho}_\text{sep}$ is optimal if and only if there exists a $\mathcal{Z}$ such that
\begin{eqnarray}
	\begin{array}{ll}
	\text{(I)} & \begin{bmatrix}
		\tilde{\varrho}_\text{sep} & 0 & 0\\
		0 & \tilde{\varrho}_\text{sep}^{\,\text{T}_1} & 0\\
		0 & 0 & \tilde{\varrho}_\text{pure}
		\end{bmatrix}
		\begin{bmatrix}
		\mathcal{Z}_1 & 0 & 0\\
		0 & \mathcal{Z}_2 & 0\\
		0 & 0 & \mathcal{Z}_3
		\end{bmatrix}=0,
		\end{array} \nonumber
\end{eqnarray}
\begin{eqnarray}
\begin{array}{ll}
	\text{(II)} & \begin{bmatrix}
		\tilde{\varrho}_\text{sep} & 0 & 0\\
		0 & \tilde{\varrho}_\text{sep}^{\,\text{T}_1} & 0\\
		0 & 0 & \tilde{\varrho}_\text{pure}
		\end{bmatrix}\geq 0,
		\end{array} \nonumber
\end{eqnarray}
\begin{eqnarray}
\begin{array}{ll}
	\text{(III)} & \begin{bmatrix}
		\mathcal{Z}_1 & 0 & 0\\
		0 & \mathcal{Z}_2 & 0\\
		0 & 0 & \mathcal{Z}_3
		\end{bmatrix}\geq 0, \quad
		\begin{array}{ll}
		\text{tr}\{F_i\mathcal{Z}\}=c_i,\\
		i=1,\ldots,16.
		\end{array}
	\end{array}\label{compslack}
\end{eqnarray}
Here, $\tilde{\varrho}_\text{sep}$, $\tilde{\varrho}_\text{pure}$, and $\mathcal{Z}$ refer to the \emph{optimal} variables. Let us digest this information. (I) is a set of three eigenstate equations from the slackness condition that determines the matrices $\mathcal{Z}_1$, $\mathcal{Z}_2$ and $\mathcal{Z}_3$. (II) is the primal constraint and simply reiterates that we have a valid decomposition in the first place. (III) is the set of dual constraints, which we will utilize to express $\mathcal{Z}_3$ in terms of $\mathcal{Z}_1$ and $\mathcal{Z}_2$. Notice that the $F_i$s are composed of blocks of $E_i$s, the 16 orthogonal basis matrices for the space of $4\times 4$ hermitian matrices. In fact, the 16 dual constraints $\text{tr}\{F_i\mathcal{Z}\}=c_i$ are really statements about the 16 components of the operator $\mathcal{Z}_1+\mathcal{Z}_2^{\text{T}_1}-\mathcal{Z}_3$ in the ``directions'' of $E_i$. Specifically, the $i\text{th}$ dual constraint reads
\begin{align}
	\text{tr}\{F_i\mathcal{Z}\} 
	&= \text{tr}\left\{\frac{1}{4}
			\begin{bmatrix}
			E_i & 0 & 0\\
			0 & E_i^{\text{T}_1} & 0\\
			0 & 0 & -E_i\end{bmatrix}
			\begin{bmatrix}
			\mathcal{Z}_1 & 0 & 0\\
			0 & \mathcal{Z}_2 & 0\\
			0 & 0 & \mathcal{Z}_3
			\end{bmatrix}\right\}\nonumber\\
	&= 	\frac{1}{4}\text{tr}\{E_i\mathcal{Z}_1\}
			+\frac{1}{4}\text{tr}\{E_i^{\text{T}_1}\mathcal{Z}_2\}
			-\frac{1}{4}\text{tr}\{E_i\mathcal{Z}_3\}\nonumber\\
	&=	\frac{1}{4}\text{tr}\{E_i(\mathcal{Z}_1+\mathcal{Z}_2^{\text{T}_1}-\mathcal{Z}_3)\}=c_i,
\end{align}
where we used the identity ${\text{tr}\{E_i^{\text{T}_1}\mathcal{Z}_2\}=\text{tr}\{E_i\mathcal{Z}_2^{\text{T}_1}\}}$. Since any hermitian operator can be written as ${H=\frac{1}{4}\sum_{i=1}^{16}{E_i\,\text{tr}\{E_iH\}}}$, we arrive at
\begin{equation}
	\mathcal{Z}_3=\mathcal{Z}_1+\mathcal{Z}_2^{\text{T}_1}+\openone_4.\label{Z3formula}
\end{equation}

We are now ready to state the Wellens--Ku\'s equations. The third block equation in (II) states, using Eq.~\eqref{Z3formula},
\begin{equation}
	(\mathcal{Z}_1+\mathcal{Z}_2^{\text{T}_1})\tilde{\varrho}_\text{pure}=-\tilde{\varrho}_\text{pure}.\label{SDPWK1}
\end{equation}
This is supplemented by the second block equation in (I), in which we carry out the replacement $\tilde{\varrho}_\text{sep}\to\rho-\tilde{\varrho}_\text{pure}$ to obtain
\begin{equation}
	(\rho-\tilde{\varrho}_\text{pure})^{\text{T}_1}\mathcal{Z}_2=0.\label{SDPWK2}
\end{equation}
Equations \eqref{SDPWK1} and \eqref{SDPWK2} are the Wellens--Ku\'s equations, which we restate here for easy reference:
\begin{eqnarray}
	\exists\alpha,\mu\geq 0 &&( \mu\rho_1+\rho_2^{\text{T}_1})\varrho_\text{pure}=-\alpha\varrho_\text{pure}, \label{WK1} \\
	&&\left( \rho-(1-\mathcal{S})\varrho_\text{pure}\right)^{\text{T}_1}\rho_2=0.\label{WK2}
\end{eqnarray}
The first block-equation in (I) states that $\tilde{\varrho}_\text{sep}\mathcal{Z}_1=0$, so $\mathcal{Z}_1$ is proportional to $\rho_1$. Therefore, Eqs.~\eqref{WK1} and \eqref{SDPWK1} are really the same equations, with the multiplicative factors $\alpha$ and $\mu$ absorbed in the normalization of $\mathcal{Z}_1$ and $\mathcal{Z}_2$. It is also clear that Eqs.~\eqref{WK2} and \eqref{SDPWK2} are the same equations, with $\rho_2$ and $\mathcal{Z}_2$ differing only by a multiplicative factor.

We remark that the barely-separable property of $\tilde{\varrho}_\text{sep}$ in the optimal LSD of $\rho$ can be derived as a consequence of this formulation. Suppose otherwise, that $\tilde{\varrho}_\text{sep}^{\text{T}_1}$ has full rank. Then we must have $\mathcal{Z}_2=0$ and Eq.~\eqref{SDPWK1} becomes $\mathcal{Z}_1\tilde{\varrho}_\text{pure}=-\tilde{\varrho}_\text{pure}$. But $\mathcal{Z}_1$ is assuredly nonnegative by (III), so $\tilde{\varrho}_\text{pure}$ must vanish, which is to say, $\rho$ was separable to begin with.

Now for a nonseparable $\rho$, $\tilde{\varrho}_\text{sep}^{\text{T}_1}$ has rank 3 so $\mathcal{Z}_2$ must be a pure state. If in addition, $\tilde{\varrho}_\text{sep}$ has full rank, $\mathcal{Z}_1$ must vanish. In this case, $\tilde{\varrho}_\text{pure}$ is the pure state associated with the negative eigenvalue of $\mathcal{Z}_2^{\text{T}_1}$, which is a Bell state \cite{sanpera98}. This is consistent with the observation made by Karnas and Lewenstein in Ref.~\cite{karnas01}.

In passing we note that $\mathcal{Z}_1,\mathcal{Z}_2$ and $\mathcal{Z}_3$ have an interesting interpretation in the language of entanglement witnesses. An entanglement witness $W$ is a hermitian operator such that $\text{tr}\{W\rho_\text{sep}\}\geq0$ for all separable states $\rho_\text{sep}$, but for some entangled state $\rho_\text{ent}$, $\text{tr}\{W\rho_\text{ent}\}<0$. The dual of the optimal LSD problem for 2-qubit systems can be written as an optimization over a constrained set of entanglement witnesses \cite{brandao05}, so that
\begin{equation}
	1-\mathcal{S}=\text{max}\big\{0,\minusmin_{{W+\openone_4\geq0}}\text{tr}\{W\rho\}\big\}.
\end{equation}
The quantity $\mathcal{Z}_1+\mathcal{Z}_2^{\text{T}_1}$ can be interpreted as the optimal entanglement witness $\mathcal{W}$ for the state $\rho$, since
\begin{equation}
	\begin{array}{l}
	\text{tr}\{\mathcal{W}\rho_\text{sep}\}\geq0 \;\;\forall \:\text{separable states}\: \rho_\text{sep}, \\
	\text{tr}\{\mathcal{W}\rho\}
	=\mathcal{S}-1<0.
	\end{array}
\end{equation}
It is optimal because ${\text{tr}\{\tilde{\varrho}_\text{sep}(\mathcal{Z}_1+\mathcal{Z}_2^{\text{T}_1})\}=0}$, so ${\mathcal{Z}_1+\mathcal{Z}_2^{\text{T}_1}}$ ``ignores'' the separable content of $\rho$, while maximally detecting the entangled part $\tilde{\varrho}_\text{pure}$ in accordance with Eq.~\eqref{SDPWK1}.

\section{Generalized Wellens--Ku\'s equations for reduced-rank states}
Since the optimal LSDs for rank-2 states are already known, it remains to characterize the rank-3 states to fully apprehend the LSD of any 2-qubit state. As a side result, Wellens and Ku\'s \cite{wellens01} generalized their equations to the reduced-rank states by treating them as the limit $x\rightarrow 0$ of the full-rank state $x\frac{1}{4}\openone_4+(1-x)\rho$. However, their approach has the implicit assumption that $\mathcal{Z}_2$ is a pure state, whereas it could also be of rank 2. The component $\mathcal{Z}_2^{\text{T}_1}$ in the optimal entanglement witness need not be the partial transpose of a pure state. As we will show, the SDP approach naturally takes care of this subtlety.

Clearly, the primal problem in the previous form is never strictly feasible if $\rho$ has rank 3. In order to utilize the complementary slackness condition, we need to modify the primal problem such that strict feasibility is restored. We denote the pure state orthogonal to $\rho$ by $\gamma$ and its concurrence by $q$. There will be two separate cases to consider: (i) $\gamma$ is entangled, and (ii) $\gamma$ is a product state.

\subsection{$\gamma$ is an entangled state}

\subsubsection{The primal problem}
We consider a parameterization in the three dimensional subspace spanned by $\rho$, which requires $3 \times 3=9$ parameters. The rank-3 projector onto the orthogonal complement of $\gamma$ is given by $P_3=\openone_4-\gamma$. We denote by $\openone_3$ its restriction to its own support. In its generic form, $\gamma$ can be written as
\begin{equation}\label{formstate}
	\gamma=\frac{1}{4}(\openone_4+p\sigma_1-p\tau_1-\sigma_1\tau_1-q\sigma_2\tau_2-q\sigma_3\tau_3),
\end{equation}
where $p=\sqrt{1-q^2}$ and $0<q\leq 1$. One can then construct an orthogonal basis $\{\Gamma_i : i=1,\ldots,9\}$ for the support of $\openone_3$, in which $\Gamma_1=\openone_3$ and the remaining $\Gamma_i$ are traceless. An explicit construction for $\{\Gamma_i\}$ can be found in \cite{englert02}. In this basis, the parameterization for the (unnormalized) rank-3 state $\tilde{\rho}_\text{sep}$ becomes $\tilde{\rho}_\text{sep}=\frac{1}{3}\vec{x}\cdot\vec{\Gamma}$, where the primal variable $\vec{x}=(\lambda,x_2,\ldots,x_9)$ is in $\mathbb{R}^9$.

One can represent the $\Gamma_i$s by $3\times 3$ matrices, but their partial transposes $\Gamma_i^{\text{T}_1}$s can be full-rank, therefore we need $4\times 4$ matrices to write them. Following the same prescription as in the full-rank case, we express the three primal constraints in block diagonal form,
\begin{equation}
		\begin{bmatrix}
		0 & 0 & 0\\
		0 & 0 & 0\\
		0 & 0 & \rho
	\end{bmatrix}+\frac{1}{3}\sum_{i=1}^9{x_i
		\begin{bmatrix}
		\Gamma_i & 0 & 0\\
		0 & \Gamma_i^{\text{T}_1} & 0\\
		0 & 0 & -\Gamma_i
	\end{bmatrix}}
	\geq 0,\label{blockconstraintrank3}
\end{equation}
where the first and third blocks are $3\times 3$ and the second block is $4\times 4$. Analogously to the full-rank case, we define $F_i=\frac{1}{3}\text{diag}(\Gamma_i, \Gamma_i^{\text{T}_1}, -\Gamma_i), i=1,\ldots,9$, and $F_0=\text{diag}(0, 0, \rho)$, so that Eq.~\eqref{blockconstraintrank3} turns into $F(\vec{x})=F_0+\sum_{i=1}^9{x_iF_i}\geq 0$. Finally, we also define $\vec{c}^{\,\text{T}}=(-1,0,\ldots,0)\in \mathbb{R}^{9}$, such that $\vec{c}^{\,\text{T}}\vec{x}=-\lambda$. With these specifications, the optimal LSD problem for rank-3 states has been cast as a SDP.

We proceed to show that this is a strictly feasible problem. The state $\rho$ has three positive eigenvalues and can be regarded as positive definite when considering only the subspace orthogonal to $\gamma$. We choose $\vec{x}^{\,\text{T}}=(\alpha,0,\ldots,0)$ where $0<\alpha /3<\text{smallest positive eigenvalue of $\rho$}$, so that
\begin{eqnarray}
	\tilde{\rho}_\text{sep}&=&\alpha\frac{1}{3}\openone_3>0, \nonumber \\
	\rho-\tilde{\rho}_\text{sep}&=&\rho-\alpha\frac{1}{3}\openone_3>0.
\end{eqnarray}
The first and third blocks of $F(\vec{x})$ are thus positive definite. For the second block, we need the fact that the eigenvalues of $\gamma^{\text{T}_1}$ are given by $\frac{1}{2}(1\pm p)$ and $\pm\frac{1}{2}q$ \cite{sanpera98}. Since $0<q\leq1$ and $0\leq p < 1$, this means that $\tilde{\rho}_\text{sep}^{\text{T}_1}=\alpha\frac{1}{3}(\openone_4-\gamma^{\text{T}_1})$ is positive definite. Let us note that $\tilde{\rho}_\text{sep}^{\text{T}_1}$ has zero eigenvalues only if $q=0$, i.e., when $\gamma$ is a product state. Thus, if we assume that $\gamma$ is \emph{not} a product state, $\tilde{\rho}_\text{sep}^{\text{T}_1}>0$ and we have strict primal feasibility. The case where $\gamma$ is a product state is treated in Sec.~\ref{section}.

\subsubsection{The dual problem}
The dual variable $Z$ is now a $10\times10$ positive semidefinite matrix, subject to nine dual constraints. Strict dual feasibility is immediate as we can choose	${Z=\text{diag}(\openone_3, \openone_4, 3\openone_3)}$, which can be easily checked to satisfy the nine dual constraints.

\subsubsection{Generalized Wellens--Ku\'s equations}
Now, having established strict primal and dual feasibility, we can invoke the complementary slackness condition \eqref{compslack}, or rather its rank-3 analog. The $i\text{th}$ dual constraint now reads
\begin{eqnarray}
\text{tr}\{F_i\mathcal{Z}\}&=&\frac{1}{3}\text{tr}\{\Gamma_i(\mathcal{Z}_1+\openone_3\mathcal{Z}_2^{\text{T}_1}\openone_3-\mathcal{Z}_3)\}\nonumber \\
	&=&c_i,\;i=1,\ldots,9.
\end{eqnarray}
Any hermitian operator orthogonal to $\gamma$ can be written as $H_\text{rank3}=\sum_{i=1}^{9}{\Gamma_i\,\text{tr}\{\Gamma_iH_\text{rank3}\}}/\text{tr}\{\Gamma_i^2\}$. Let us repeat here that both $\mathcal{Z}_2$ and $\mathcal{Z}_2^{\text{T}_1}$ have support in the total Hilbert space. To avoid inconsistency in the notation, let us define $\mathcal{Z}_{2||}^{\text{T}_1}$, the restriction of the projection $P_3{Z}_2^{\text{T}_1}P_3$ to its own support. We then arrive at $\mathcal{Z}_3=\mathcal{Z}_1+\mathcal{Z}_{2||}^{\text{T}_1}+\openone_3$. The third block equation in (I) of Eq.~\eqref{compslack} then states that $(\mathcal{Z}_1+\mathcal{Z}_{2||}^{\text{T}_1}+\openone_3)\tilde{\varrho}_\text{pure}=0,$ and since $\tilde{\varrho}_\text{pure}$ resides in the subspace that $\openone_3$ projects onto,
\begin{equation}
	(\mathcal{Z}_1+\mathcal{Z}_{2||}^{\text{T}_1})\tilde{\varrho}_\text{pure}=-\tilde{\varrho}_\text{pure},\label{SDPWK1rank3}
\end{equation}
and as before, this is supplemented by the eigenstate equation for $\mathcal{Z}_2$,
\begin{equation}
	(\rho-\tilde{\varrho}_\text{pure})^{\text{T}_1}\mathcal{Z}_2=0.\label{SDPWK2rank3}
\end{equation}
Equations \eqref{SDPWK1rank3} and \eqref{SDPWK2rank3} are the generalization of the Wellens--Ku\'s equations to the rank-3 case where the orthogonal state is entangled. These are almost identical to the original equations, the subtle difference being that not only $Z_2$, but also $Z_{2||}^{\text{T}_1}$, the projection of its partial transpose onto the support of $\rho$, are now relevant. Similarly to the full-rank case, one can define $\mathcal{Z}_1+\mathcal{Z}_{2||}^{\text{T}_1}$ as the optimal entanglement witness for the state $\rho$.

\subsection{$\gamma$ is a product state}\label{section}
\subsubsection{The primal problem}
A little more care is needed if $\rho$ is orthogonal to a pure product state $\gamma=\frac{1}{2}(\openone_4+\sigma_1)\frac{1}{2}(\openone_4-\tau_1)$, the $q=0$ version of Eq.~\eqref{formstate}. In this case, since $\tilde{\rho}_\text{sep}$ is separable and orthogonal to $\gamma$, $\tilde{\rho}_\text{sep}^{\text{T}_1}$ and $\gamma^{\text{T}_1}$ must be orthogonal too. The separability of $\tilde{\rho}_\text{sep}$ then requires: (i) the positivity of $\tilde{\rho}_\text{sep}^{\text{T}_1}$, and (ii) the orthogonality of $\tilde{\rho}_\text{sep}^{\text{T}_1}$ and $\gamma^{\text{T}_1}$. Only two of the nine $\Gamma_i$s do \emph{not} obey $\Gamma_i^{\text{T}_1}\gamma^{\text{T}_1}=0$. These are ${\Gamma_8=\frac{1}{2}(\sigma_2\tau_2-\sigma_3\tau_3)}$ and $\Gamma_9=\frac{1}{2}(\sigma_2\tau_3+\sigma_3\tau_2)$. Furthermore, there exists a proportionality relation between the products $\Gamma_8^{\text{T}_1}\gamma^{\text{T}_1}$ and $\Gamma_9^{\text{T}_1}\gamma^{\text{T}_1}$, inasmuch as $\Gamma_8^{\text{T}_1}\gamma^{\text{T}_1}=\frac{1}{2}(\Gamma_8^{\text{T}_1}+\text{i}\Gamma_9^{\text{T}_1})=\text{i}\Gamma_9^{\text{T}_1}\gamma^{\text{T}_1}$. Constraint (ii) then reads
\begin{equation}
	\tilde{\rho}_\text{sep}^{\text{T}_1}\gamma^{\text{T}_1}=\frac{1}{6}(x_8-\text{i}x_9)(\Gamma_8^{\text{T}_1}+\text{i}\Gamma_9^{\text{T}_1})=0.
\end{equation}
Since $x_8$ and $x_9$ are real, they must vanish and we have the parameterization $\tilde{\rho}_\text{sep}=\frac{1}{3}\sum_{i=1}^7{x_i\Gamma_i}$. Consequently, the primal objective is now $\vec{c}^\text{T}\vec{x}$ with $\vec{x}\in\mathbb{R}^7$. The same choice of $\tilde{\rho}_\text{sep}=\alpha\frac{1}{3}\openone_3$ shows that this modified primal problem is strictly feasible. 

\subsubsection{The dual problem}
One can also verify, in the now familiar manner, that $Z=\text{diag}(\openone_3, \openone_3^{\text{T}_1}, 3\openone_3)$ is a strictly feasible point for the modified dual problem. 

\subsubsection{Generalized Wellens--Ku\'s equations}
The seven dual constraints lead to ${\mathcal{Z}_1+\mathcal{Z}_{2||}^{\text{T}_1}-\mathcal{Z}_3+a\Gamma_8+b\Gamma_9=-\openone_3}$, where $a$ and $b$ are some real coefficients. We then arrive at another pair of generalized Wellens--Ku\'s equations,
\begin{align}	(\mathcal{Z}_1+\mathcal{Z}_{2||}^{\text{T}_1}+a\Gamma_8+b\Gamma_9)\tilde{\varrho}_\text{pure}&=-\tilde{\varrho}_\text{pure}, \label{SDPWKspecial1}\\
	(\rho-\tilde{\varrho}_\text{pure})^{\text{T}_1}\mathcal{Z}_2&=0,\label{SDPWKspecial2}
\end{align}
which are necessary and sufficient for optimality. Note that $\mathcal{Z}_2$ lies in the support of $\openone_3^{\text{T}_1}$ while $\mathcal{Z}_2^{\text{T}_1}$ can have support in the total Hilbert space since $\mathcal{Z}_2$ is not separable. The term in parentheses in Eq.~\eqref{SDPWKspecial1} is the optimal entanglement witness for $\rho$. In contrast with the earlier cases, nonpositivity is provided by the combination $\mathcal{Z}_{2||}^{\text{T}_1}+a\Gamma_8+b\Gamma_9$. 

\subsubsection{$\tilde{\varrho}_{\rm sep}$ has rank 3}
In the full-rank case, when the separable part is full-rank, the nonseparable part is maximally entangled. A similar property exists for rank-3 states. In three dimensions, the analog of the full-rank case is a rank-3 state orthogonal to a pure product state. Note that the pure state has to be a product state to ensure that all the relevant positive operators remain of rank 3 under partial transposition. If $\tilde{\varrho}_\text{sep}$ has rank 3, the optimal decomposition can be obtained analytically. In this case, ${\mathcal{Z}_1=\mathcal{Z}_2=0}$ and Eq.~\eqref{SDPWKspecial1} reduces to ${(a\Gamma_8+b\Gamma_9) \tilde{\varrho}_\text{pure}=-\tilde{\varrho}_\text{pure}}$. In the magic basis, which is a basis of Bell states, the nonzero matrix elements of $a\Gamma_8+b\Gamma_9$ appear as
\begin{equation}
	a\Gamma_8+b\Gamma_9\widehat{=}
	\begin{bmatrix}
	 a & b\\
	 b & -a
	\end{bmatrix},
\end{equation}
where the two basis states are $\ket{\phi^+}=\frac{1}{\sqrt{2}}(\ket{00}+\ket{11})$ and $\ket{\psi^+}=\frac{\text{i}}{\sqrt{2}}(\ket{01}+\ket{10})$. Equation~\eqref{SDPWKspecial1} imposes that this matrix has an eigenvalue $-1$. Putting this requirement into its characteristic equation leads to the relation $a^2+b^2=1$, and an angular parameterization, $a=\cos{\theta}, b=\sin{\theta}$, can be used. The corresponding eigenstate is nondegenerate, and can hence be identified with $\tilde{\varrho}_\text{pure}$. Explicitly, we have
\begin{equation}
	\ket{\varrho_\text{pure}}=\cos{\frac{\theta}{2}}\ket{\phi^+} - \sin{\frac{\theta}{2}}\ket{\psi^+},
\end{equation}
which is maximally entangled.

Now, $\tilde{\varrho}_\text{sep}$ has no components along $\Gamma_8$ and $\Gamma_9$, so one must have $\text{tr}\{\Gamma_i (\rho-\tilde{\varrho}_\text{pure})\}=0 \;\; \textrm{for} \; i=8,9$. These turn out to provide a simple set of equations for the unknowns $\theta$ and $\mathcal{S}$:
\begin{equation}
	\begin{array}{rl}
	\text{tr}\{\Gamma_8\rho\}&=(\mathcal{S}-1)\cos{\theta},\\
	\text{tr}\{\Gamma_9\rho\}&=(\mathcal{S}-1)\sin{\theta}.
	\end{array}\label{specialsimultaneous}
\end{equation}
Therefore, we obtain $\mathcal{S}=1-\sqrt{(\text{tr}\{\Gamma_8\rho\})^2+(\text{tr}\{\Gamma_9\rho\})^2}$. The solution to Eq.~\eqref{specialsimultaneous} then gives us $\tilde{\varrho}_\text{pure}$ and ${\tilde{\varrho}_\text{sep}=\rho-\tilde{\varrho}_\text{pure}}$ in the optimal LSD of $\rho$. 

In general, one can assume that $\tilde{\varrho}_\text{sep}$ has rank 3 and use the above result to determine the optimal $\tilde{\varrho}_\text{sep}$ and $\tilde{\varrho}_\text{pure}$. It is however necessary to check if the deduced $\tilde{\varrho}_\text{sep}$ is indeed of rank 3 and separable. If the verification fails, $\tilde{\varrho}_\text{sep}$ has rank 2 and one has to solve the generalized Wellens--Ku\'s equations given in Eqs.~\eqref{SDPWKspecial1} -- \eqref{SDPWKspecial2}.

\section{Conclusion}
We have demonstrated that the problem of finding the optimal LSD of a 2-qubit state is a SDP. Indeed, the Peres-Horodecki criterion has permitted us to advantageously rephrase a separability constraint as a positivity constraint. We have shown that both the primal and the associated dual programs are strictly feasible, leading us to necessary and sufficient optimality conditions for LSD. In particular we have derived the original Wellens--Ku\'s equations for full-rank states in a simple and natural way. Moreover we have generalized them to rank-3 states. We have also described the link between the dual SDP variables and entanglement witnesses. Finally, many efficient algorithms for solving SDPs are available, allowing one to handle this problem numerically. Because the Peres--Horodecki criterion is also necessary and sufficient for composite systems of dimensions $2\times 3$, it might be possible to extend the SDP formulation to this case.


\begin{acknowledgments}
TGC wishes to thank the National University of Singapore for granting him an undergraduate scholarship under which this study was carried out. Centre for Quantum Technologies is a Research Centre of Excellence funded by Ministry of Education and National Research Foundation of Singapore.
\end{acknowledgments}

\end{document}